\documentclass[aps,superscriptaddress,preprint]{revtex4}

\usepackage{latexsym,graphicx,epsfig,psfrag,fancybox,xspace}
\usepackage{amssymb,amsmath,amsxtra,amsfonts}
\usepackage{bm}

\usepackage{array}
\usepackage{booktabs}
\usepackage{empheq}
\usepackage{float}
\usepackage{xcolor}
\usepackage{wrapfig}
\usepackage{multirow}
\usepackage{longtable}

\newcommand{\bea}{\begin{eqnarray}}
\newcommand{\ena}{\end{eqnarray}}
\newcommand{\be}{\begin{equation}}
\newcommand{\en}{\end{equation}}
\newcommand{\nn}{\nonumber\\}
\newcommand{\ed}{\end{document}} 
\newcommand{\Tr}{\mbox{\rm{tr}}}

\newcommand{\Br}{\ensuremath{\mathcal{B}}\xspace}

\begin{document}
  
\title{$B \to K^{(\ast)} \nu \bar{\nu}$ in covariant confined quark model} 

\author{Aidos Issadykov}
\email{issadykov@jinr.ru}
\affiliation{Bogoliubov Laboratory of Theoretical Physics, 
Joint Institute for Nuclear Research, 141980 Dubna, Russia}
\affiliation{Institute of Nuclear Physics, 
  Ministry of Energy of the Republic of Kazakhstan, 050032 Almaty, Kazakhstan}
\author{Mikhail A. Ivanov}
\email{ivanovm@theor.jinr.ru }
\affiliation{Bogoliubov Laboratory of Theoretical Physics, 
Joint Institute for Nuclear Research, 141980 Dubna, Russia}

\begin{abstract}
 
We study the $B \to K^{(\ast)} \nu \bar{\nu}$ decays within
the Standard Model (SM) by using the relevant transition form factors
obtained from the covariant confined quark model (CCQM)  developed by us.
The   $B \to K$  and $B\to K^\ast$ transition form factors are calculated
in the full kinematic $q^2$ range. The branching fractions are then calculated.
It is shown that our results are in an agreement with those  obtained
in other theoretical approaches.  Currently, the Babar and Belle 
collaborations provide us by the upper limits at 90\% confidence limit.
The obtained bounds are roughly an order of magnitude larger than the SM
predictions. This should stimulate experimental collaborations to set up
experiments that allow one to obtain more accurate branching values, which is
quite achievable on the updated LHCb and Belle machines. If the discrepancies
between theory and experiment are confirmed, this will open up opportunities
for constructing models with new particles and interactions leading to an
extension of the SM.

\end{abstract}
\maketitle

\section{Introduction}
\label{sec:intro}

The rare weak decays
$B \to K^{(\ast)} \nu \bar{\nu}$ ($K^{(*)} = K\,\, \text{or}\,\, K^ \ast$)
proceed via a $b-s$ flavor changing neutral current (FCNC), i.e. at
loop level in the Standard Model (SM). Such transitions are significantly
suppressed in such a way that the branching fractions are of $10^{-6}$ order.
Nevertheless, the decays with a lepton pair in the final state
are observed and studied by BaBar and  Belle collaborations.
The decays with di-neutriono in the final state have been only observed
and the upper limits established~\cite{Belle:2007vmd,BaBar:2008wiw,BaBar:2010oqg,Belle:2013tnz,BaBar:2013npw,Belle:2017oht,Belle-II:2021rof}.

Theoretically, the rare $B$-meson decays to final states, containing a pair
of neutrinos are among the most ``clean'' processes with neutral
currents that change the flavors of quarks. Since the neutrino is electrically
neutral, the factorization of hadron and lepton currents in this decay is
exact, in contrast to other decays of $B$ mesons. For this reason, precise
measurement of the $B \to K^{(*)} \nu \bar{\nu}$ processes should allow one to
extract the $B \to K^{(*)}$ transition form factors with high accuracy.
Another advantage of such decays is that they are free of charmonium
resonance contributions. A detailed analysis of these decays has been performed
in the series of publications by A.~Buras and his collaborators~\cite{Altmannshofer:2009ma,Buras:2014fpa,Buras:2021nns,Buras:2022wpw}, recently the lattice
QCD determination of $B-K$ transition form factors have been used to determine
SM differential branching fractions for
$B\to K\nu\bar\nu$~decay~\cite{Parrott:2022dnu}. There are some other
theoretical approaches to this activity, see, for instance,
\cite{Kamenik:2009kc,Colangelo:1996ay,Melikhov:1997zu,Faessler:2002ut,Wang:2012ab,Du:2015tda,Browder:2021hbl}. 

The $B \to K^{(*)} \nu \bar{\nu}$ processes can also be associated with
other decays of the $B$-meson, which proceed through the formation of some
exotic state that decays into a pair of neutrinos. Such signals are
interesting in the context of the problem of searching for dark matter and
may allow us to investigate the relationship between the Standard Model and
the so-called dark sector of the Universe~\cite{Kamenik:2011vy}.
The interplay between dineutrino modes and semileptonic rare
B-decays has been elaborated in Ref.~\cite{Bause:2021cna}.

Current experimental bounds on the $B\to K\nu\bar\nu$ branching fraction
are roughly an order of magnitude larger than the SM predictions. Thus,
increasing the accuracy of measuring $B \to K^{(*)} \nu \bar{\nu}$
processes is an extremely important task.

In this paper we focus on these decays within the SM framework using
the relevant form factors calculated in the
covariant constituent quark model (CCQM) previously developed by us.

\section{A general information on the decay of $B \to K^{(\ast)} \nu \bar{\nu}$}

The effective Hamiltonian for $b\to s\nu\bar\nu$~transitions in the SM
can be written in the form, see, for instance~Ref.~\cite{Buras:2014fpa}
and other references therein
\be
{\mathcal H}_{\rm eff}^{\rm SM} =
\frac {G_F}{\sqrt2}  V_{tb} \, V^*_{ts}
\Big[\frac {\alpha_{em}}{2\pi} \frac {X_t}{\sin^2\Theta_W}\Big]
(\bar s\, O_\mu\, b) (\bar\nu O^\mu \nu) + h.c.
\en
Here $G_F$  is the Fermi constant,  $V_{tb}$ and  $V^\ast_{ts}$
are the matrix elements of the Cabibbo-Kobayashi-Maskawa (CKM) matrix,
$\alpha_{em}$ is the fine structure constant calculated at the
electroweak scale of $Z$-boson mass,  $\theta_W$ is the Weinberg angle,
$O_\mu=\gamma_\mu(1-\gamma_5)$ is the weak Dirac matrix with the left
chirality. The Wilson coefficient $X_t$ is known with a high accuracy,
including NLO QCD corrections~\cite{Misiak:1999yg,Buchalla:1998ba} and
two-loop electroweak contributions~\cite{Brod:2010hi}, resulting in
\be
X_t=1.469 \pm 0.017.
\en

Matrix elements $\langle K^{(\ast)} | \bar s O^\mu b) | B \rangle$ can be
written in terms of scalar functions --form factors depending on 
the transferred momentum squared 
\bea
\langle K(p_2)\,|\,\bar s\, O^{\,\mu}\, b\,| B(p_1)\rangle
&=&
F_+(q^2)\, P^{\,\mu} + F_-(q^2)\, q^{\,\mu},
\label{eq:PP'}\\[1.5ex]
\langle K^\ast(p_2,\epsilon_2)\,|\,\bar s\, O^{\,\mu}\,b\, |\,B(p_1)\rangle
&=&
\frac{\epsilon^{\,\dagger}_{2\,\alpha}}{m_1+m_2}\,
\Big( - g^{\mu\alpha}\,Pq\,A_0(q^2) + P^{\,\mu}\,P^{\,\alpha}\,A_+(q^2)
\nn
&+& q^{\,\mu}\,P^{\,\alpha}\,A_-(q^2)
+ i\,\varepsilon^{\mu\alpha P q}\,V(q^2)\Big).
\label{eq:PV}
\ena
Here $P =p_1+p_2$, \,$q=p_1-p_2$, and $\epsilon_2$ is the polarization vector
of the $K^\ast$ meson, so $\epsilon_2^\dagger\cdot p_2 = 0$ .
The abbreviation
$\varepsilon^{\mu\alpha P q} = \varepsilon^{\mu\alpha \nu\delta} P_\nu q_\delta$
is adopted, where the absolutely antisymmetric Levi-Civita tensor is
defined as $\varepsilon^{0123} = -\varepsilon_{0123} = -1$.
The particles are on the mass shell, i.e.
$p_1^2=m_1^2=m_B^2$ and $p_2^2=m_2^2=m_{K^{(\ast)}}^2$.

It should be noted that the calculation of form factors requires methods
outside the framework of perturbation theory, such as, for example, various
quark models, QCD sum rules, lattice QCD, etc.
Taking into account the above definitions of form factors,
the differential branching fraction is  written as follows:
\begin{align}
\frac{ d \Br (B\to K^{(\ast)}+\nu\bar\nu)}{dq^2} &=
3\tau_{B} \frac{ (G_F \lambda_t \alpha_{em})^2 }{3(2\pi)^5}
\Big(\frac{X_t}{\sin^2\theta_W}\Big)^2
\nn
&\times \frac{|{\bf p_2}|}{4m_1^2}\,
\Big( {\tilde H}_+^2 + {\tilde H}_-^2 + {\tilde H}_0^2 \Big).
\label{eq:BR}
\end{align}
The factor 3 at the beginning of the formula comes from the summation over
the neutrino flavors: $\nu_e$, $\nu_\mu$ and $\nu_\tau$. In the Standard Model,
they are assumed to be massless and therefore do not interfere with each other.
Further, $\tau_{B}$ is the B-meson lifetime,
$\lambda_t\equiv| V_{tb}V^\ast_{ts} |$,
$|{\bf p_2}|=\lambda^{1/2}(m_1^2,m_2^2,q^2)/2m_1 $ is the  daughter meson momentum
in the rest frame of the initial $B$~meson.
Scaled helicity amplitudes ${\tilde H}$
are related to the helicity amplitudes defined in \cite{Ivanov:2015tru}
as ${\tilde H} = \sqrt{q^2} H$. As a result, we have\\[1.2ex]
\noindent
{\boldmath $B\to K$ \,\,\bf transition:}
\be
{\tilde H}_\pm = 0,
\qquad
{\tilde H}_0 = 2\,m_1\,|{\bf p_2}|\,F_+.
\label{eq:hel_PP}
\en
\noindent
{\boldmath$B\to K^\ast$ \,\,\bf transition:}
\bea
{\tilde H}_\pm &=&
\frac{\sqrt{q^2}}{m_1+m_2}
\Big(-Pq\, A_0 \pm 2\,m_1\,|{\bf p_2}|\, V \Big),
\nn[1.2ex]
{\tilde H}_0 &=&
\frac{1}{m_1+m_2}\frac{1}{2\,m_2}
\Big(-Pq\,(Pq - q^2)\, A_0 + 4\,m_1^2\,|{\bf p_2}|^2\, A_+\Big).
\label{eq:hel_PV}
\ena

The scaling is done in order to avoid uncertainties at the point $q^2 = 0$ in
numerical calculations. We note that in these decays the physical range of
the variable $q^2$ is $0\le q^2 \le q^2_{\rm max}=(m_1-m_2)^2$.
It is interesting to note that in the case of a $B\to K$ transition,
differential branching fraction behaves like $|{\bf p_2}|^3$ near the end point
of the spectrum
\be
\frac{ d \Br (B\to K+\nu\bar\nu)}{dq^2} =
\tau_{B} \frac{ (G_F \lambda_t \alpha_{em})^2 }{(2\pi)^5}
\Big(\frac{X_t}{\sin^2\theta_W}\Big)^2 \cdot
|{\bf p_2}|^3\,|F_+(q^2)|^2,
\label{eq:BR-BK}
\en
and disappears as $(q^2-q^2_{\rm max})^{3/2}$.
Note that in contrast to $B\to K^{(\ast)} \ell^+\ell^-$ decays,
the isospin asymmetries of the $B\to K^{(\ast)} \nu\bar\nu$ vanish identically,
so the branching fractions of the $B^0$ and $B^\pm$ decays only differ
due to the lifetime difference~\cite{Buras:2014fpa}:
$\tau_{B^\pm}=1.638\pm 0.004$~ps and $\tau_{B^0}=1.519\pm0.004$~ps.
In this paper we will consider the charged B-meson only.

\section{Form factors in CCQM}

In the CCQM the nonlocal quark interpolating currents are used to describe
the internal structure of a hadron
\bea
 J_M(x) &=& \int\!\! dx_1 \!\!\int\!\! dx_2\, F_M (x;x_1,x_2)\cdot
 \bar q^a_{f_1}(x_1)\, \Gamma_M \,q^a_{f_2}(x_2)
 \hspace{1.4cm} \text{Meson}
\nn
 J_B(x) &=& \int\!\! dx_1 \!\!\int\!\! dx_2 \!\!\int\!\! dx_3\,
 F_B (x;x_1,x_2,x_3)
\hspace{3.0cm}  \text{Baryon}
\nn
&&
\times\, \Gamma_1 \, q^{a_1}_{f_1}(x_1) \,
 \Big[ 
\varepsilon^{a_1a_2a_3}  q^{T\,a_2}_{f_2}(x_2) C \, \Gamma_2 \,  q^{a_3}_{f_3}(x_3)
              \Big]
\nn
 J_T(x) &=& 
\int\!\! dx_1\ldots\int\!\! dx_4\, 
 F_T (x;x_1,\ldots,x_4)
\hspace{2.8cm} \text{ Tetraquark}
\nn
&&
\times\,
 \Big[ 
\varepsilon^{a_1a_2c} q_{f_1}^{T\,a_1}(x_1)\, C\Gamma_1\, q_{f_2}^{a_2}(x_2)
                  \Big]
\cdot 
 \Big[
\varepsilon^{a_3a_4c} \bar q_{f_3}^{T\,a_3}(x_3)\, \Gamma_2 C\, \bar q_{f_4}^{a_4}(x_4)
               \Big]
\ena
The vertex functions $F_H$ are chosen in the translational invariant form
\be
F_H(x;x_1,\ldots,x_n) = \delta\left(x-\sum\limits_{i=1}^n w_i x_i\right)
\Phi_H\left(\sum\limits_{i<j} (x_i-x_j)^2\right),
\en
where $w_i = m_i/\sum\limits_{j=1}^n m_j$.
The function $\Phi_H$ is taken Gaussian in such a way that its
Fourier-transform decreases  quite rapidly in the Euclidean direction
and provides the ultraviolet convergence of the Feynman diagrams.

\begin{figure}[H]
\centering  
\includegraphics[scale=0.5]{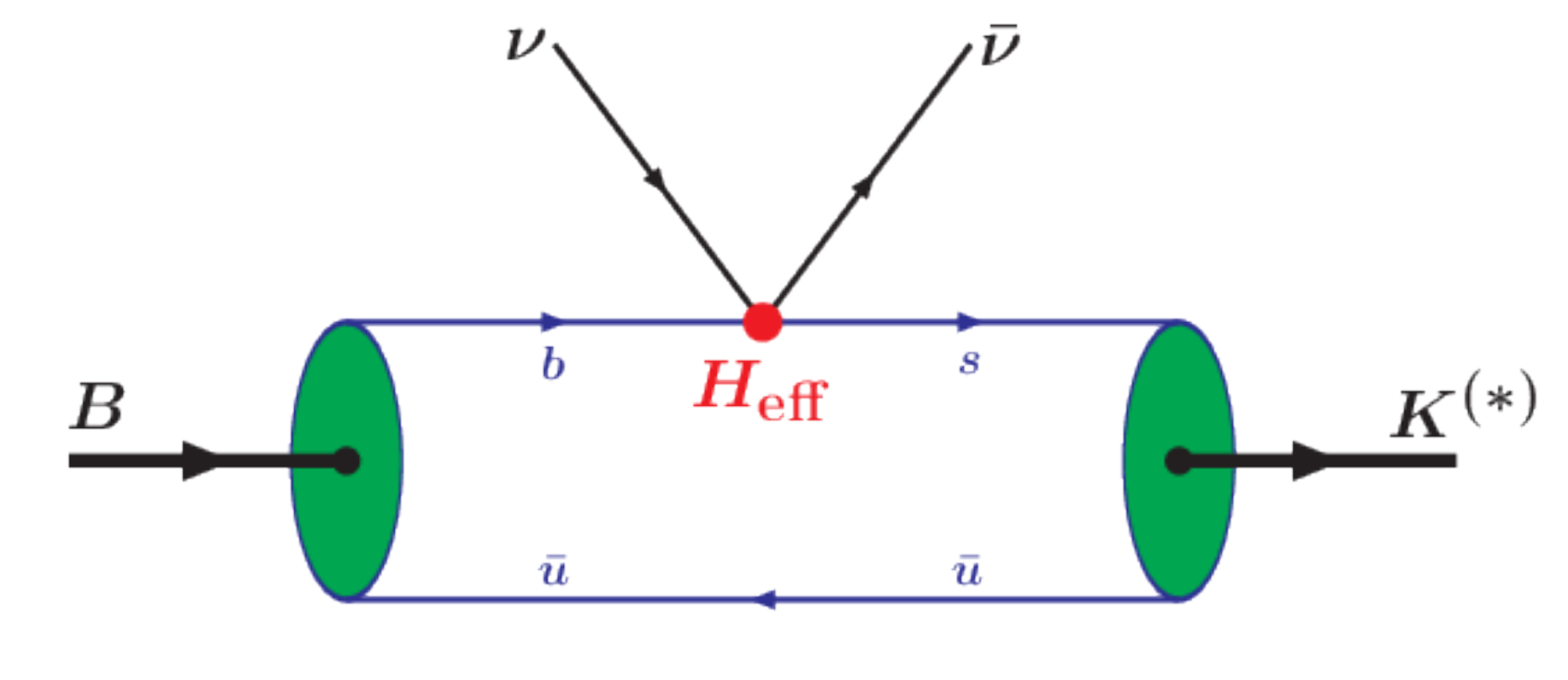}
\caption{Diagram describing the decay of $B \to K^{(\ast)} \nu\bar\nu$
in the CCQM.}
\label{fig:B-K}
\end{figure}

\begin{figure}[H]
\begin{tabular}{lr}
\includegraphics[scale=0.55]{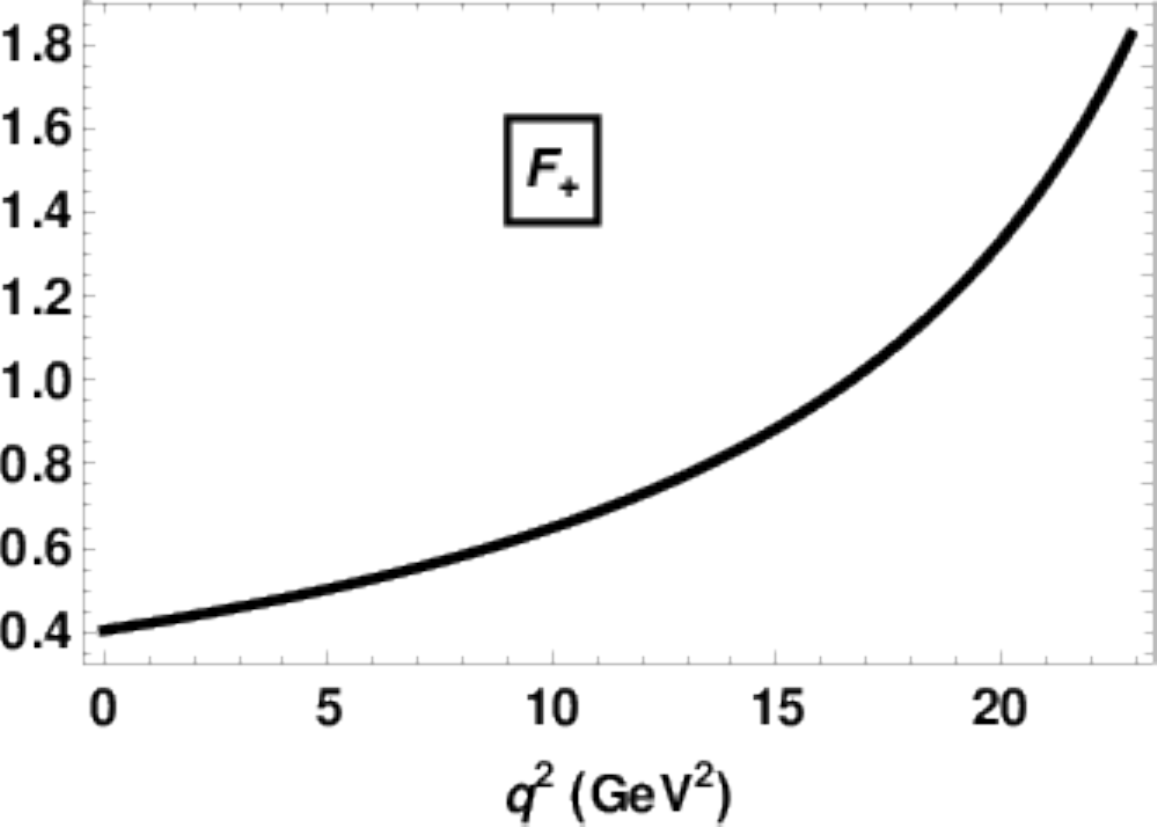} &
\includegraphics[scale=0.55]{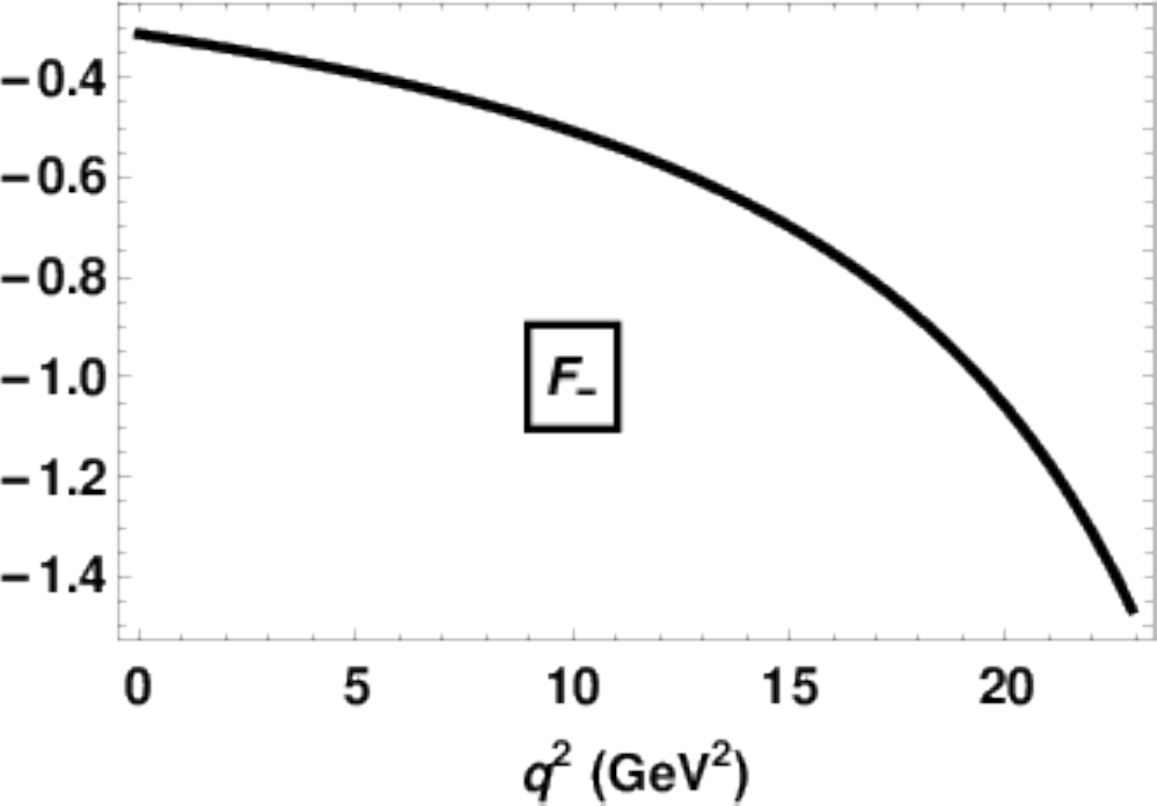}\\
\end{tabular}
\caption{Form factors of $B \to K$ transition.}
\label{fig:formfactors-BK}
\end{figure}

\begin{figure}[H]
\begin{tabular}{lr}
\includegraphics[scale=0.55]{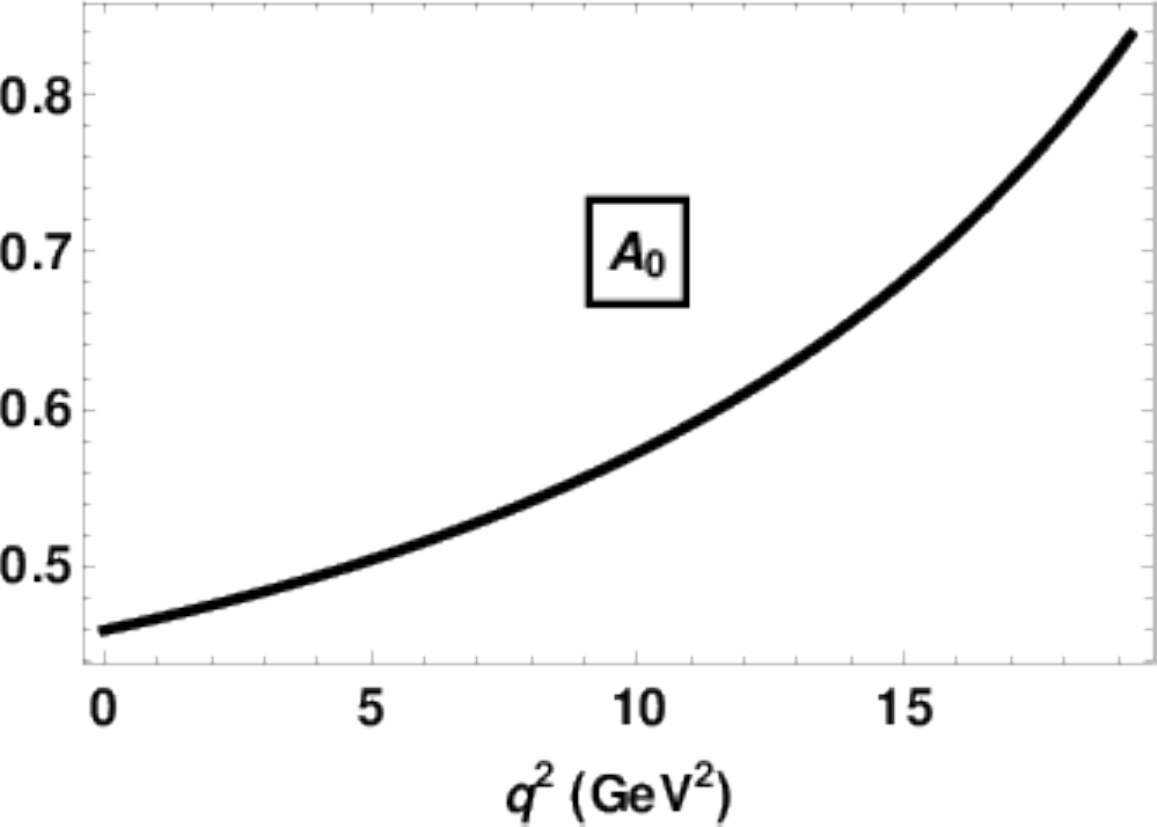}&
\includegraphics[scale=0.55]{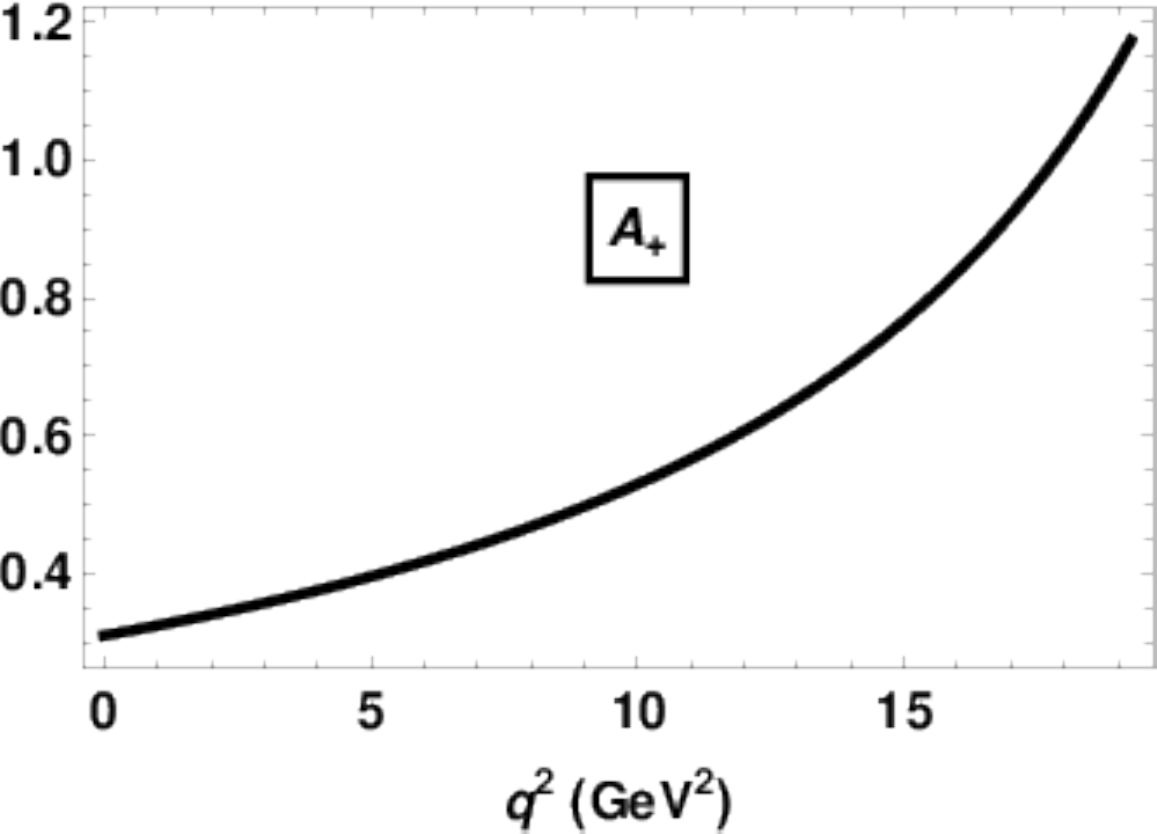}\\
\includegraphics[scale=0.55]{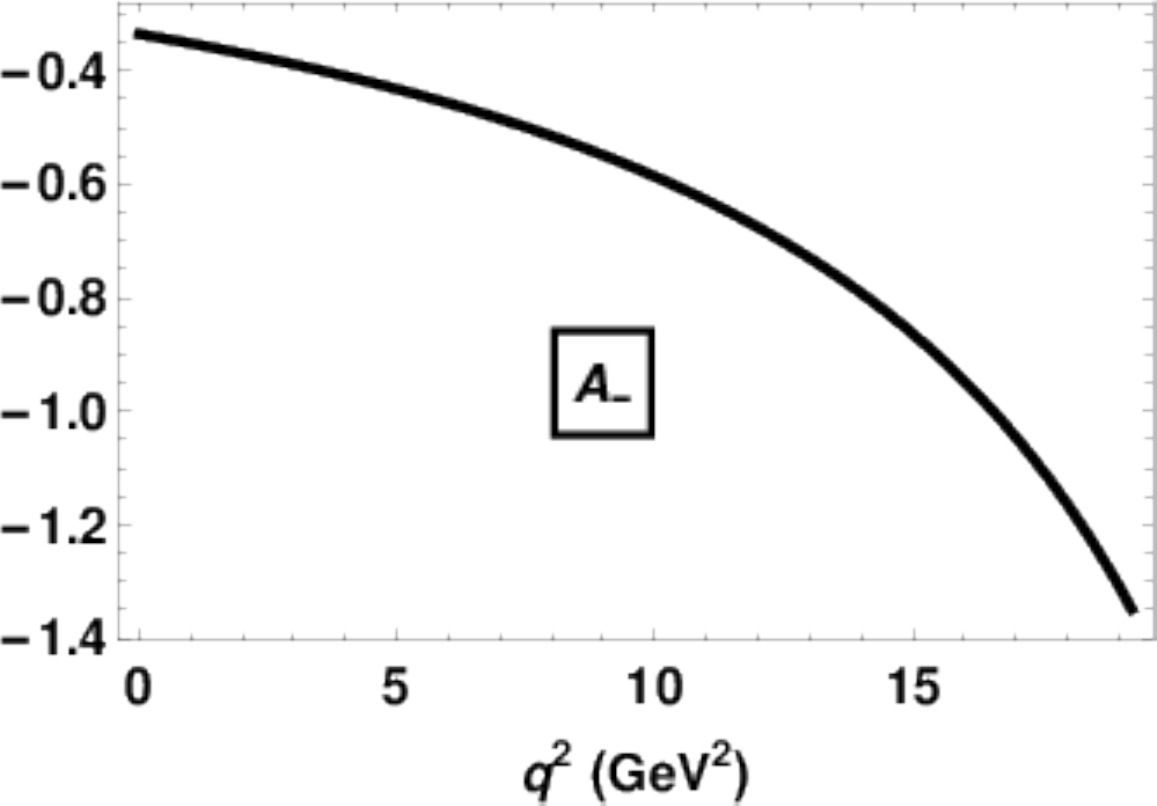}&
\includegraphics[scale=0.55]{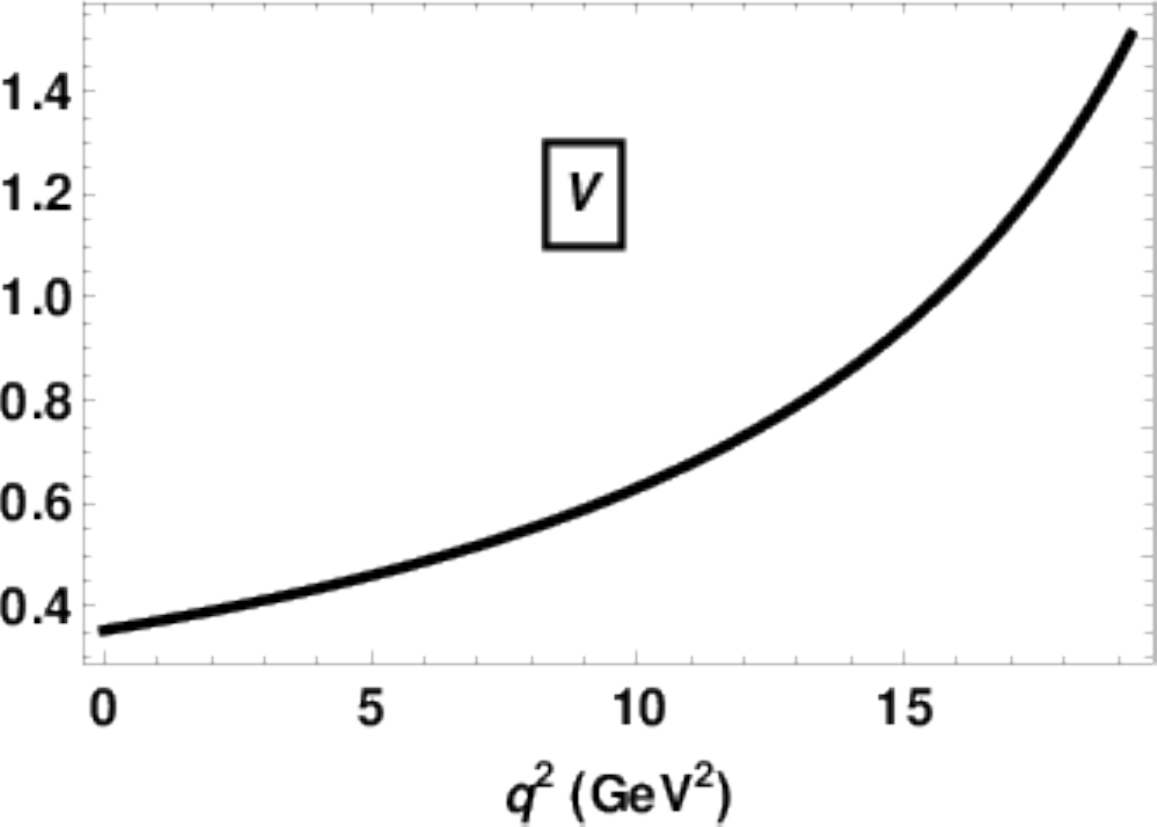}\\
\end{tabular}
\caption{Form factors of $B \to K^\ast$ transition.}
\label{fig:formfactors-BKstar}
\end{figure}

The matrix elements of the rare decay $B \to K^{(\ast)} \nu\bar\nu$ are
described by the diagram shown in Fig.~\ref{fig:B-K}.
The invariant matrix elements of the weak current between
the initial and final meson states are written down
\bea
&&
\langle K(p_2)\,|\,\bar s\, O^{\,\mu}\, b\,| B(p_1) \rangle \,=\,
\nn[1.2ex]
&=&
N_c\, g_B\,g_K\!\!  \int\!\! \frac{d^4k}{ (2\pi)^4 i}\, 
\widetilde\Phi_B\Big(-(k+w_{13} p_1)^2\Big)\,
\widetilde\Phi_K\Big(-(k+w_{23} p_2)^2\Big)
\nn
&\times&
\Tr \biggl[
O^{\,\mu}\, S_1(k+p_1)\, \gamma^5\, S_3(k)\, \gamma^5\, S_2(k+p_2) 
\biggr]
\nn[1.2ex]
& = & F_+(q^2)\, P^{\,\mu} + F_-(q^2)\, q^{\,\mu},\quad 
\label{eq:model_PP}\\[1.5ex]
&&
\langle K^\ast(p_2,\epsilon_2)\,|\,\bar s\, O^{\,\mu}\,b\, |\,B(p_1) \rangle 
\,=\,
\nn[1.2ex]
&=&
N_c\, g_B\,g_{K^\ast} \!\! \int\!\! \frac{d^4k}{ (2\pi)^4 i}\, 
\widetilde\Phi_B\Big(-(k+w_{13} p_1)^2\Big)\,
\widetilde\Phi_{K^\ast}\Big(-(k+w_{23}p_2)^2\Big)
\nn
&\times&
\Tr \biggl[ 
O^{\,\mu} \,S_1(k+p_1)\,\gamma^5\, S_3(k) \not\!\epsilon_2^{\,\,\dagger} \,
S_2(k+p_2)\, \biggr]
\nn[1.2ex]
& = &
\frac{\epsilon^{\,\dagger}_{2\,\alpha}}{m_1+m_2}\,
\left( - g^{\mu\alpha}\,Pq\,A_0(q^2) + P^{\,\mu}\,P^{\,\alpha}\,A_+(q^2) 
+ q^{\,\mu}\,P^{\,\alpha}\,A_-(q^2) \right.
\nn
&& \hspace*{3cm} \left.
+ i\,\varepsilon^{\mu\alpha P q}\,V(q^2)\right).
\label{eq:model_PV}
\ena
We introduce 2-index notation for the reduced masses
$w_{ij} = m_{q_j}/(m_{q_i}+m_{q_j})$  $(i,j=1,2,3)$,
so that $w_{ij}+w_{ji}=1$. In our case we have $q_1=b$, $q_2=s$ and $q_3=u$.
Using the technique described in our previous papers, see, for example,
\cite{Branz:2009cd,Ivanov:2011aa,Issadykov:2018myx,Tran:2018kuv,Ivanov:2016qtw},
the final expressions for the form factors are represented as two-fold parametric integrals.
Numerical results for the form factors $F(q^2)$ can be approximated
by the dipole formula with high accuracy
\be
F(q^2)=\frac{F(0)}{1 - a s + b s^2}, \quad s=\frac{q^2}{m_1^2}. 
\label{eq:DPP}
\en
The relative error of such an approximation is less than $1\%$ in the entire kinematic region.
The parameters $F(0),a,b$  are given by Eq.~(\ref{eq:ff_param}).
\be
\begin{array}{c|rr|rrrr}
	& \quad F_+ \quad & \quad  F_- \quad & \quad A_0 \quad  & \quad A_+ \quad & 
	\quad A_- \quad & \quad V \quad \\
	\hline
		F(0) &  0.40   & -0.31 &  0.46 & 0.31  & -0.34 & 0.35
		\\
		a    &  1.13   &  1.16 &  0.44 & 1.25  &  1.31 & 1.35
		\\ 
		b    &  0.22  & 0.24   & -0.31 & 0.27  &  0.32 & 0.34
		\\ 
		\hline
\end{array}
\label{eq:ff_param}
\en
The behavior of the form factors $F_+,\,F_-$ and $A_0,\,A_+,\,A_-,V$
in the entire kinematic region $0\le q^2 \le q^2_{\rm max}$,
where $q^2_{\rm max}=(m_1-m_2)^2$, are shown in
Fig.~\ref{fig:formfactors-BK} and \ref{fig:formfactors-BKstar}.

\section{Numerical results}

\begin{figure}[H]
\begin{tabular}{lr}
\includegraphics[scale=0.55]{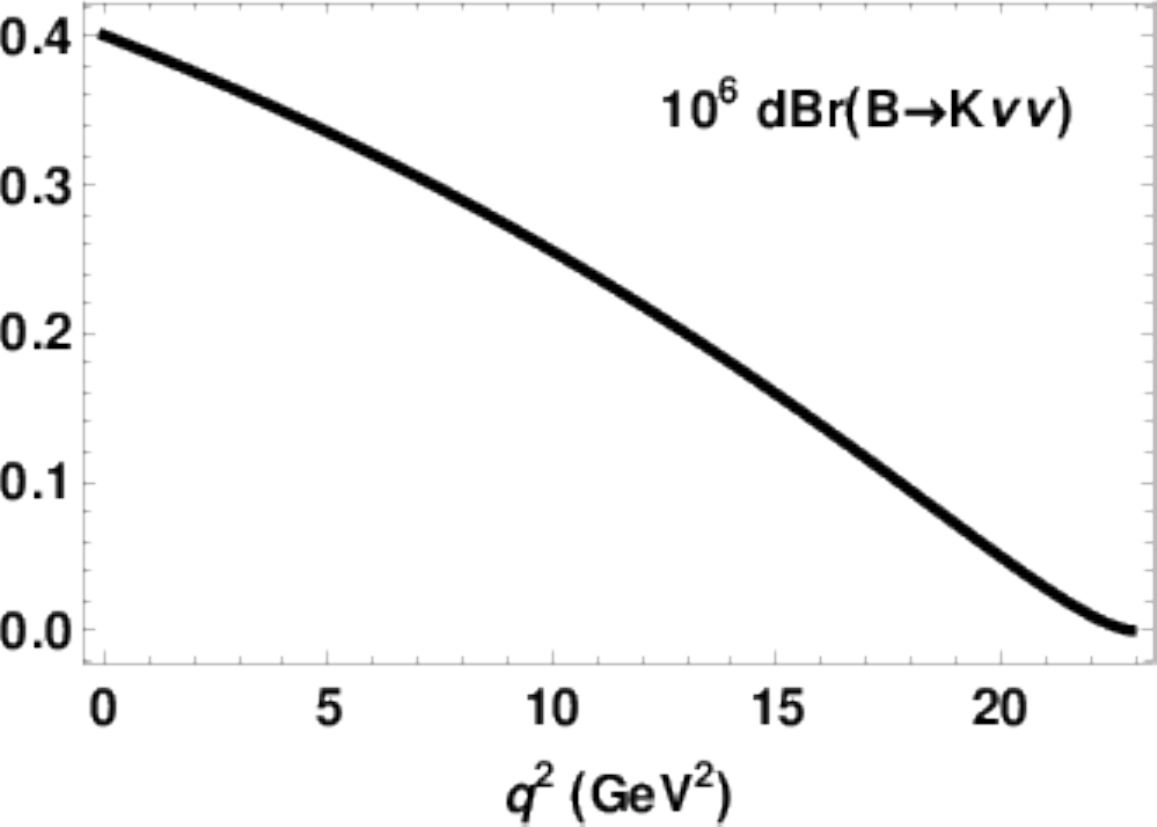} &
\includegraphics[scale=0.55]{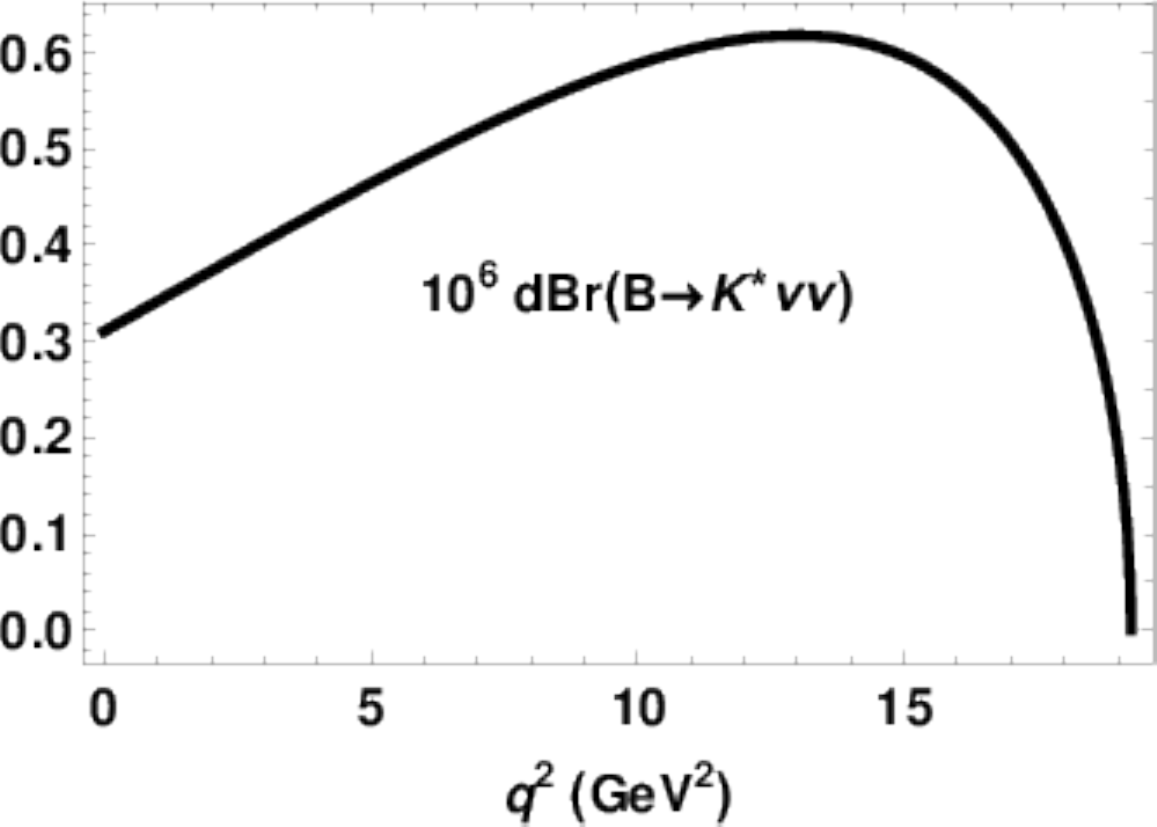}\\
\end{tabular}
\caption{Behavior of differential branchings of $B \to K^{(\ast)}\nu\bar\nu$ decays over the entire kinematic region of the squared momentum transfer.}
\label{fig:branching}
\end{figure}

In this section, we present numerical results for both differential and
integrated branching fractions of the $B \to K^{(\ast)}\nu\bar\nu$ decays.
In Table~\ref{tab:input} we present numerical values of parameters
required for calculations~\cite{ParticleDataGroup:2022pth}.
\begin{table}[H]
\caption{Input parameters used in calculations.}
\label{tab:input}
\begin{center}
\def\arraystretch{1.2}
\begin{tabular}{cc||cc}
\hline
$m_{B^+}$ & 5279.25(26) \ \ MeV & $\tau_{B^+}$ & 1.638(4)\,\,ps \\
$m_{K^+}$ & 493.677(13) \ MeV & \ \ $\sin^2\theta_W$ & 0.23126(5)  \\
$m_{K^{\ast\,+}}$ & 891.67(26) \ \ MeV & $\alpha_{em}^{-1}$ & 127.925(16)  \\
$G_F$  & $1.166\times 10^{-5}$\,\,GeV$^{-2}$ \ \ & $\lambda_t$ & 0.0401(10)  \\
\hline
\end{tabular}
\end{center}
\end{table}
The behavior of differential branching fractions
of $B \to K^{(\ast)}\nu\bar\nu$ decays
in the entire kinematic region of the  momentum transfer squared
are shown in Fig.~\ref{fig:branching}. It should be noted that in the case
of a $B\to K$ transition, differential branching fraction behaves as
$(q^2-q^2_{\rm max})^{3/2}$ near the end point of the spectrum,
in contrast to $B \to K^\ast$ transition, where the behavior is
$(q^2-q^2_{\rm max})^{1/2}$.

Table~\ref{tab:BR_tot} shows the results of calculations of the total
branching fractions of $B \to K^{(\ast)}\nu\bar\nu$, obtained in our work, and,
for comparison, the results of the experimental measurements and
theoretical predictions obtained in variiuos approaches.
We estimate the errors of the results of our model at 15 $\%$.
%
\begin{table}[H]
\caption{Summary of experimental and theoretical results for the branching
  fractions $\mathcal{B}(B\to K^{(\ast)}\nu\bar{\nu})$. Experimental upper
  limits are given at 90\% confidence level.}
\label{tab:BR_tot}  
  \begin{center}
    \def\arraystretch{1.2}
\begin{tabular}{cc|cc}
  \hline
  \multicolumn{2}{c}{$10^6\Br(B^+\to K^+\nu\bar\nu)$} &
  \multicolumn{2}{c}{$10^6\Br(B\to K^{\ast}\nu\bar\nu)$} \\
\hline\hline
Exp. (90\% CL) & Ref.\ \ & Exp. (90\% CL) & Ref. \ \  \\[1.2ex]
\hline
$<14$  & \cite{Belle:2007vmd} &  $<140$ &  \cite{Belle:2007vmd} \\

$<13$  & \cite{BaBar:2010oqg} &  $<80$  & \cite{BaBar:2008wiw}  \\

$<55$  & \cite{Belle:2013tnz} &  $<40$ & \cite{Belle:2013tnz} \\

$<16$  & \cite{BaBar:2013npw} &  $<64$ & \cite{BaBar:2013npw} \\

$<19$  & \cite{Belle:2017oht} & $<61$  & \cite{Belle:2017oht} \\

$<41$  & \cite{Belle-II:2021rof}   &  & \\[1.2ex]
\hline
Theory & Ref. & Theory & Ref. \\
\hline

 2.4(0.6) & \cite{Colangelo:1996ay} & 5.1(0.8) & \cite{Colangelo:1996ay} \\

 5.2(1.1) & \cite{Melikhov:1997zu}  & 13(5) & \cite{Melikhov:1997zu}  \\

 4.19(42) & \cite{Faessler:2002ut} &&\\

 4.5(0.7) & \cite{Altmannshofer:2009ma} & 6.8$^{+1.0}_{-1.1}$ &
 \cite{Altmannshofer:2009ma} \\

 5.1(0.8) & \cite{Kamenik:2009kc}& 8.4(1.4) & \cite{Kamenik:2009kc}  \\

 4.23(56) & \cite{Bause:2021cna} & 8.93(1.07) & \cite{Bause:2021cna} \\

 $4.4^{+1.4}_{-1.1}$ & \cite{Wang:2012ab}&&\\

 4.0(0.5) & \cite{Buras:2014fpa} & 9.2(1.0) & \cite{Buras:2014fpa} \\
 
 4.94(52) & \cite{Du:2015tda}&&\\

 4.45(62) &\cite{Buras:2021nns}& 9.70(92)  &\cite{Buras:2021nns} \\
 
 4.65(62) &\cite{Buras:2022wpw}& 10.13(92) & \cite{Buras:2022wpw} \\

 5.67(38) & \cite{Parrott:2022dnu} &&\\

 4.96(0.74) & this work & 9.57(1.43)& this work \\[1.2ex]
 \hline
\end{tabular}
\end{center}
\end{table}

 
 \section{Summary}
 \label{sec:summary}
We have provided a thorough analysis of the semileptonic decays
$B\to K^{(\ast)}\nu\bar\nu$ in the SM by using the relevant form factors
obtained from our covariant confined quark model.
The decay branching fractions  have been calculated and the following
results obtained
\begin{align}
  \Br(B^+ \to K^+\nu\bar\nu)        & = 4.96(0.74) \times 10^{-6},
  \nn
\Br(B^+ \to K^{\ast\,+}\nu\bar\nu) &  = 9.57(1.43) \times 10^{-6}.
\nonumber
\end{align}
The results are in an agreement with those  obtained
in other theoretical approaches. 
Currently, the Babar and Belle collaboratiions provide us by
the upper limits at 90\% confidence limit.
The obtained bounds are roughly an order of magnitude larger than the SM
predictions.
This stimulates experimental collaborations to set up
experiments that allow one to obtain more accurate branching values, which is
quite achievable on the updated LHCb and Belle machines. If the discrepancies
between theory and experiment are confirmed, this will open up opportunities
for constructing models with new particles and interactions leading to an
extension of the Standard Model.

\begin{acknowledgments}
We thank Egor Tretyakov who participated at the early steps of this work.
 This research has been funded by the Science Committee of the Ministry
 of Education and Science of the Republic of Kazakhstan (Grant No. AP09057862).
\end{acknowledgments}



\ed